\documentclass{nature}
\usepackage{times}
\usepackage{xcolor}
\usepackage{graphicx}
\usepackage{amsmath}
\usepackage{physics}
\usepackage{gensymb}
\usepackage{color}
\usepackage{url}
\usepackage{tabularx}
\usepackage[final]{changes}
\makeatletter
\@namedef{Changes@AuthorColor}{red}
\colorlet{Changes@Color}{red}
\makeatother
\definechangesauthor[name={Added}, color=red]{new}
\definechangesauthor[name={Deleted}, color=blue]{old}

\title{Device-system Co-design of Photonic Neuromorphic Processor using Reinforcement Learning}

\author{Yingheng Tang,$^{1,2}$ Princess Tara Zamani,$^{1}$ Ruiyang Chen,$^{1} $Jianzhu Ma,$^{3,4}$ Minghao Qi,$^{2}$ Cunxi Yu,$^{1,\ast}$ and Weilu Gao$^{1,\ast}$}

\begin{document}

\maketitle

\begin{affiliations}
 \item Department of Electrical and Computer Engineering, The University of Utah, Salt Lake City, UT 84112, USA
 \item Department of Electrical and Computer Engineering, Purdue University, West Lafayette, IN 47907, USA
 \item Institute of Artificial Intelligence, Peking University, Beijing, China
 \item Beijing Institute of General Artificial Intelligence, Beijing, China
\end{affiliations}
\noindent $^\ast$To whom correspondence should be addressed; E-mail: cunxi.yu@utah.edu; weilu.gao@utah.edu.

\pagebreak
\begin{abstract}

The incorporation of high-performance optoelectronic devices into photonic neuromorphic processors can substantially accelerate computationally intensive operations in machine learning (ML) algorithms. However, the conventional device design wisdom is disconnected with system optimization. We report a device-system co-design methodology to optimize a free-space optical general matrix multiplication (GEMM) hardware accelerator by engineering a spatially reconfigurable array made from chalcogenide phase change materials. With a highly-parallelized hardware emulator constructed based on experimental information, we demonstrate the design of unit device by optimizing GEMM calculation accuracy via reinforcement learning, including deep Q-learning neural network, Bayesian optimization, and their cascaded approach, which show a clear correlation between system performance metrics and physical device specifications. Furthermore, we employ physics-aware training approaches to deploy optimized hardware to the tasks of image classification, materials discovery, and a closed-loop design of optical ML accelerators. The demonstrated framework offers insights into the co-design of optoelectronic devices and systems with reduced human-supervision and domain-knowledge barriers. 

\end{abstract}

\pagebreak

\section*{Introduction}
\vspace{-20pt}
Fast and efficient processing of machine learning (ML) algorithms has become crucial to many aspects of modern technologies, such as image processing and computer vision\cite{LeCunEtAl2015N,GoodfellowEtAl2016}, the prediction and discovery of materials and molecules\cite{ButlerEtAl2018N,SeniorEtAl2020N}, and the chip design of artificial intelligence accelerators\cite{MirhoseiniEtAl2021N}. One of the most computation-intense operations in ML algorithms is general matrix multiplication (GEMM), which requires substantial computation and memory resources and consumes significant amount of energy. There have been extensive efforts devoted to accelerating these operations by building energy-efficient specific electronic hardware\cite{SzeEtAl2017PI}, including neuromorphic processing architecture implemented through nonvolatile resistive memories\cite{ChuaEtAl1971ITCT,WangEtAl2017NM,BoybatEtAl2018NC,HuEtAl2018AM,LiEtAl2018JPAP}. However, the power consumption and density of integrated electronic circuits have started to hit a bottleneck of processing algorithms with trillions of arithmetic operations.

Emerging efforts on leveraging fundamentally different particles, photons, show appealing promise for building high-throughput and power-efficient hardware to accelerate GEMM operations, thanks to the extreme optical parallelism and low static power consumption. Both two-dimensional silicon photonic integrated circuits (PICs)\cite{ShenEtAl2017NP,HarrisEtAl2018O} and three-dimensional (3D) free-space optical systems\cite{LinEtAl2018S,HamerlyEtAl2019PR,ZhouEtAl2021NP,WangEtAl2022NC} have been demonstrated. Furthermore, the incorporation of novel optical materials provide unprecedented functionalities and performance of optoelectronic devices used in optical ML hardware accelerators. For example, nonvolatile chalcogenide phase change materials (PCMs) enable the in-memory computing (i.e., co-located data storage and processing) in PICs, which further reduces energy consumption\cite{FeldmannEtAl2021N}. However, the current efforts of designing individual optoelectronic components based on novel materials is decoupled with the target performance of optical ML accelerators. Thus, it is elusive to translate system specifications into the engineering aims on material and device levels.

Here, we demonstrate a device-system co-design methodology to optimize a 3D free-space optical GEMM (O-GEMM) accelerator by engineering a spatially reconfigurable array (SRA) based on germanium-antimony-tellurium (GST) PCMs through reinforcement learning (RL). We develop a highly-parallelized hardware emulator for such optimization, which incorporates materials optical properties, a physical device calculation engine, and an architecture simulation engine. The high-level GEMM calculation generated from this hardware emulator can be directly employed in various ML algorithms. We utilize RL algorithms of deep Q-learning neural networks, Bayesian optimization, and a hybrid approach to maximize the reward function defined from the GEMM calculation accuracy by adjusting actions corresponding to device and materials parameters. Despite without including the knowledge of optoelectronic devices in SRAs in RL algorithms, we clearly observe that the high-accuracy calculation of O-GEMM is strongly correlated to the large transmission, modulation depth, and generally small thickness of individual optoelectronic devices. This suggests that our methodology relaxes the requirement of human-supervised design and reduces domain-knowledge barriers for non-expert device designers. Furthermore, we demonstrate the deployment of O-GEMM hardware through the approaches involving physics-aware training to different ML algorithms for various applications, including the image classification in three standard datasets, the prediction of 2D material properties, and the new design of O-GEMM hardware by implementing RL algorithms using O-GEMM hardware. 

\section*{Results}
\begin{figure}
    \includegraphics[width=0.9\textwidth]{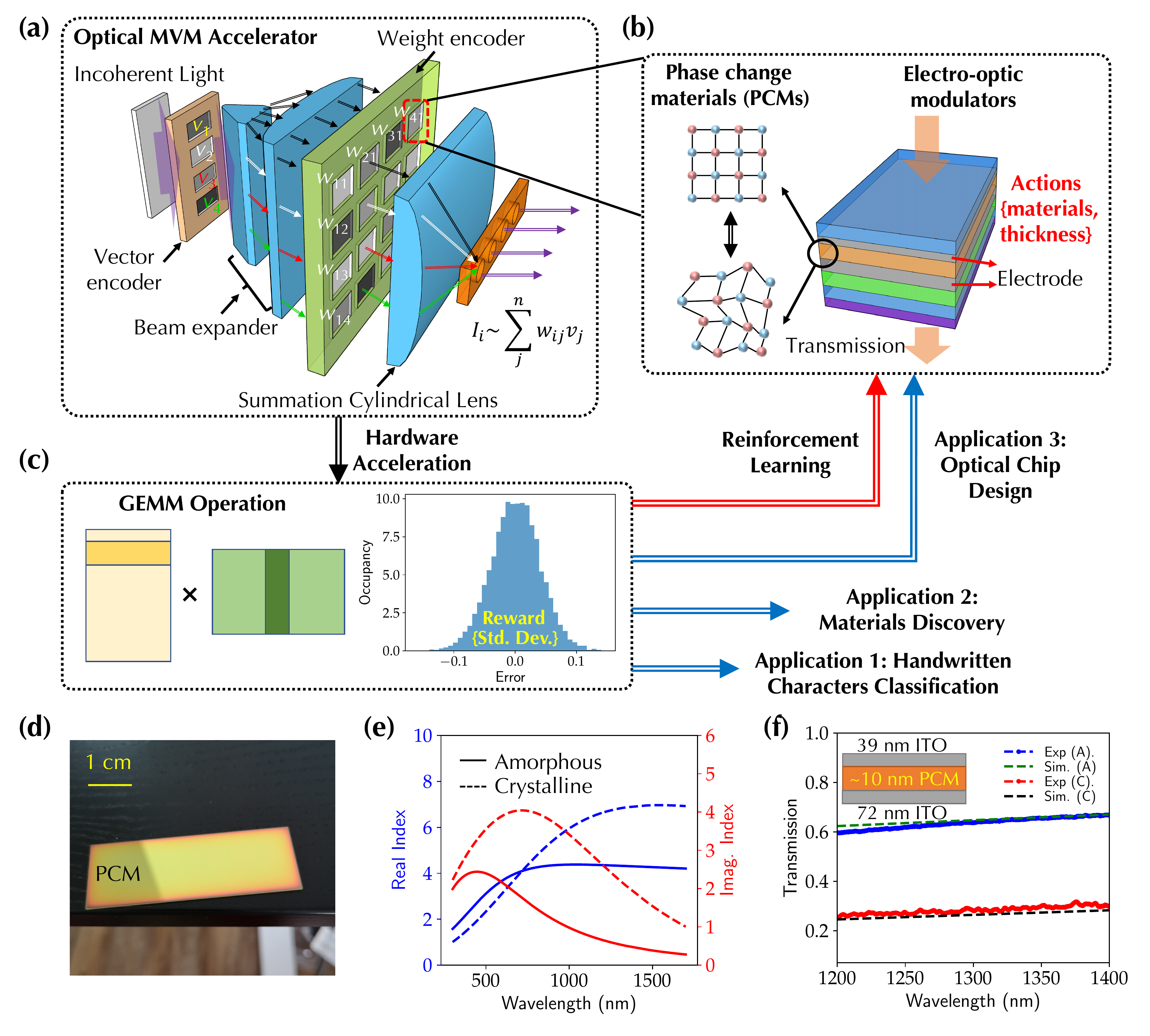}
    \caption{\label{fig:arch} \textbf{Device-system co-design, applications, and demonstrations of optical general matrix multiplication (GEMM) hardware.} (a)\,An illustration of optical matrix-vector multiplication (MVM) hardware accelerator. 
    (b)\,An illustration of a weight reconfigurable unit consisting a stack of multiple thin films with non-volatile phase change materials (PCMs). The choice of materials and their corresponding thicknesses become the actions in reinforcement learning (RL) algorithms. (c)\,GEMM calculation based on the optical MVM accelerator. The standard deviation of the errors in the calculations of randomly generated input matrices is used as the reward function in RL algorithms. The optical-GEMM is used in multiple machine learning applications, such handwritten characters classification, nanomaterials discovery, and the chip design of optical GEMM hardware. (d)\,An experimental deposition of a wafer-scale GST-PCM thin film on a silicon substrate. (e)\,The complex-valued refractive indices of GST-PCM at crystalline and amorphous phases measured by visible and near-infrared ellipsometry. (f)\,Experimentally measured and simulated transmission spectra of a fabricated multilayer stack consisting of a $39$-nm thick ITO layer, a $10$-nm thick GST-PCM layer, and a $72$-nm thick ITO layer. The phase of as-fabricated GST is amorphous and after heating under $200^{\circ}$C it becomes crystalline phase.}
  \end{figure}

\vspace{-20pt}
Figure\,\ref{fig:arch}a illustrates a 3D free-space optical accelerator for matrix-vector multiplication (MVM) and MVM-based GEMM operations when performing ML inference. In this system, an uniform and collimated incoherent light is incident onto a 1D SRA that is the vector encoder. The information in input vector is physically represented by the electrically controllable transmittance of individual modulators in the 1D SRA. In order to accommodate high-speed input data for inference, the modulators in the vector encoder need to be fast enough (e.g., $>$\,GHz). Possible implementations of such high-speed modulators can be based on metal plasmonics\cite{CaiEtAl2009NL}, graphene\cite{QiuEtAl2015OL}, free-carrier in $p-n$ junctions\cite{QiuEtAl2012SR}, and Pockels effect\cite{PengEtAl2019OE}. The vector encoder design is not the focus of this work and we assume an arbitrary monotonic function of light transmittance with respect to external electrical stimulus. 

Figure\,\ref{fig:arch}a shows an example of the multiplication of a $4\times4$ matrix \textbf{W} with a $4\times1$ vector $\vectorarrow{\textbf{v}}$. The output light from a vector unit modulator (e.g., $v_1$) is transformed to a uniformly distributed line profile and mapped onto a row of weight encoders (e.g., $w_{1j}, j=1,2,3,4$) through a beam expander, which can be experimentally realized through a combination of a Powell lens and a convex cylindrical lens. Each reconfigurable unit in the weight encoder independently regulates the transmittance, so that the intensity of output light is the multiplication of the intensity of input light and the transmittance of corresponding unit. The information of input weight matrix is physically represented by the electrically controllable transmittance of the individual units in the 2D SRA. The output light from a column of reconfigurable units transmits through a convex cylindrical lens for summation. The curved face of this summation cylindrical lens is orthogonal to that in the beam expander. The focused light is collected by an array of photodetectors. The generated photocurrent from each photodetector in the array, $I_i$, is proportional to $\Sigma_{j}v_jw_{ij}$; see Ref.\,\cite{GaoEtAl2021APR}. In order to implement negative values based on non-negative physical quantities, such as transmittance and and photocurrent, each $v_i$ and $w_{ij}$ are represented as a difference of two positive values\cite{GaoEtAl2021APR,GoodmanEtAl2005PEC}; see \emph{Methods} for more details.

In the inference of ML models, the weight matrix is generally fixed and reconfigured occasionally when the task changes. Thus, it is most desirable for the weight SRA to possess nearly zero static energy consumption together with reconfigurability capability for the best energy efficiency. In this regard, nonvolatile PCMs become an ideal material platform. After the reconfigurability of PCMs through either electrical or optical excitation, the optical properties of PCMs are preserved even after removing the stimulus\cite{HosseiniEtAl2014N,WuttigEtAl2017NP}. As a consequence, the static energy consumption of PCM-based reconfigurable unit is nearly zero, and the memory-like properties of PCMs further eliminates the energy consumption associated with the data transfer between computing and memory units. 

Figure\,\ref{fig:arch}b shows a multilayer thin film stack of PCM-based reconfigurable unit that is able to spatially adjust the transmittance of incident light. The drastic structural change of PCMs between crystalline and amorphous phases features a large refractive index change, which is on the order of unity. In practice, such phase transition can be induced by flowing currents through PCMs to heat them up. The electrode material we choose is indium tin oxide (ITO), which is relatively transparent in the wavelength of interest at $1.3\,\mu$m. We also include dielectric materials with high refractive index contrast, $n_\mathrm{SiO_2} = 1.5$ and $n_\mathrm{Si_3N_4} = 3.0$, and metallic materials, gold (Au) and aluminum (Al). The periodic structures consisting of alternating layers of SiO$_2$ and Si$_3$N$_4$ or a pair of thin metallic layers can in principle form a Fabry-Per\'ot cavity to significantly enhance the electric field and the light-matter interaction in the PCM thin layer. However, such domain knowledge is not included in the optimization of a stack of thin layers with a material list consisting of PCM, ITO, SiO$_2$, Si$_3$N$_4$, Au, and Al and varying thickness; see \emph{Methods} for more details. Specifically, we pre-define a stack of $6$ layers and in each layer we can choose any material in the material list with an arbitrary thickness in a broad range. The combination of materials and corresponding thicknesses becomes the action that is taken in RL algorithms. The transmittance of the selected stack can be experimentally measured through a home-built near-infrared spectrometer, and can be calculated using transfer matrix formalism in the hardware emulator as described below; see \emph{Methods} for more details on the experimental measurement setup and \emph{Supplementary Information Section 1} for a detailed description of transfer matrix formalism. 

Figure\,\ref{fig:arch}c describes the reward function that is used in RL algorithms. The transmittance of each unit in the vector encoder and that of each PCM-based reconfigurable unit in the weight encoder physically represent the mathematical elements of input vector and matrix, and the obtained photocurrent from photodetectors corresponds to an element of output vector. The MVM operation performed by the optical hardware can be constructed to perform GEMM through block multiplications; see \emph{Supplementary Information Section 2} for more mathematical details. The O-GEMM accelerator performance is evaluated using the calculation accuracy of performing GEMM. Specifically, we randomly generate all elements of two input matrices with each element in the range of $[-1,1]$. The first element of output matrix is used as a benchmark of calculation accuracy. There is a difference between the calculated element value and the expected value calculated from a standard computer and their difference defines the calculation error. After the GEMM calculations of $10000$ randomly generated pairs of input matrices, one obtained representative calculation error histogram is plotted in Fig.\,\ref{fig:arch}c. The reward function is defined based on the standard deviation of the calculation error; see \emph{Methods} for a detailed mathematical description.

We construct a highly-parallelized hardware emulator that consists of materials parameters, device structures, physical transfer-matrix calculation engines, optical architecture emulation, and a GEMM calculation user interface. We assume that GST-PCM can achieve 30 states\cite{LiEtAl2019O} and there is a shot noise added on the photodetector output; see \emph{Supplementary Information Section 2} for more details on the hardware emulator structure. We employ RL algorithms (red arrow in Fig.\,\ref{fig:arch}) to close the loop of the optimization of materials, devices, and systems. Specifically, the materials and device structure parameters become the action and the system performance of GEMM calculation is the reward function. The optimized O-GEMM accelerator offers the functionality of performing GEMM, which are the core functions of a handful of ML algorithms. We demonstrate a few examples of utilizing the developed optical accelerator in the applications of image classification and materials discovery. The prediction accuracy in these applications is improved through physics-aware training process\cite{WrightEtAl2022N}. Furthermore, we show that the developed O-GEMM hardware accelerator can be further employed in the RL algorithms to accelerate the chip design of another optical accelerator. 

Figure\,\ref{fig:arch}d shows an experimentally deposited large-scale ($>\mathrm{cm}^2$) GST-PCM film on a silicon substrate through RF sputtering. The thickness of the GST film can be controlled by adjusting deposition time. Figure\,\ref{fig:arch}e displays the complex refractive indices of fabricated GST films when they are in amorphous and crystalline phases. The as-deposited GST film is amorphous and the corresponding complex refractive index was experimentally measured through ellipsometry. The obtained films were heated up at $\sim200^{\circ}\mathrm{C}$ for $30\,$min and the films were converted to crystalline phase. A substantial refractive index change can be observed in Fig.\,\ref{fig:arch}e. Figure\,\ref{fig:arch}f displays a multilayer structure consisting of two transparent electrode ITO layers and a thin layer of PCM material. The thickness of top ITO layer is $72\,$nm, the PCM layer is $\sim10\,$nm, and the bottom ITO layer is $39\,$nm. The experimentally measured transmittance spectra in Fig.\,\ref{fig:arch}f are fully consistent with the calculated spectra from the hardware emulator, which is constructed based on the transfer matrix formalism. The ITO index is extracted from Ref.\,\cite{OzcarizEtAl2020S} and the optical constants of other materials used in the emulator are either experimentally measured or obtained from literatures; see \emph{Methods} for more experimental details of thin film deposition and characterization. 

Figure\,\ref{fig:RL}a summarized three employed RL algorithms to optimize GST-based SRA through maximizing the system reward function. The first method (Path 1) is based on deep Q-learning neural network (DQNN), the second method (Path 2) is based on Bayesian optimization, and the third strategy (Path 3) is to combine these two methods in a cascaded manner. A legitimate reconfigurable unit in the weight encoder requires the existence of GST material in the stack. In practice, the GST material can be electrically controlled only when two ITO electrodes are adjacent to the GST layer. As a result, in all three algorithms, a thin film GST layer must exist and the adjacent layers have to be ITO. Figure\,\ref{fig:RL}b describes a training curve of the DQNN method; see \emph{Methods} for the detailed parameters of the DQNN model. The accumulated average reward calculated based on GEMM calculation accuracy increases with training iterations when adjusting device structures, suggesting a successful device optimization for optimal O-GEMM system performance. In the validation process, a set of randomly generated devices were input into the trained DQNN model. Figure\,\ref{fig:RL}c shows that the reward increases with respect to iterations, indicating again the successful device optimization based on system metrics. Simultaneously, we extract the maximum transmittance ($T_\mathrm{max}$), transmittance modulation ($T_\mathrm{diff}$), and the total thickness of the device at each iteration, as shown in Figs.\,\ref{fig:RL}d -- f, respectively. The shaded area represents the variations of tested devices in the DQNN model. Clearly, the increase of the reward, thus the decrease of O-GEMM calculation error, is synchronized with the increase of $T_\mathrm{max}$ and $T_\mathrm{diff}$ and the decrease of total thickness, although the only optimization target in the algorithm is the reward from O-GEMM calculation accuracy. From the physical point of view, the large $T_\mathrm{max}$ and $T_\mathrm{diff}$ in the SRA indicate a strong robustness against the noise from detector and finite-bit-quantization\cite{GaoEtAl2021APR}. In addition, thin devices generally have large transmittance, while the selection of materials also matters. The correlation of the reward function with physical quantities of devices suggests that the DQNN model actually captures the essential device specifications for designing optimal O-GEMM accelerators. 

In addition to the DQNN model, we also utilize Bayesian optimization to optimize the device structure; see \emph{Methods} for details. Similar to the DQNN model, the reward increases with iterations as shown in Fig.\,\ref{fig:RL}g, which is also correlated with the increase of $T_\mathrm{max}$ and $T_\mathrm{diff}$ and the decrease of total thickness; see Figs.\,\ref{fig:RL}h -- j. The Bayesian optimization requires substantial amount of training time to have the probabilistic model to generate high-reward devices in large chances. The trained DQNN model can help accelerate the Bayesian process and make the process more deterministic. Specifically, we cascade the Bayesian optimizer with the DQNN model, where the sampled devices from a partially optimized Bayesian optimizer become the input of the DQNN model; see \emph{Methods} for details. After fast iterations through the trained DQNN model, similar to the first two methods, the reward also increases as shown in Fig.\,\ref{fig:RL}k. The reward increase is also correlated with the increase of $T_\mathrm{max}$ and $T_\mathrm{diff}$; see Figs.\,\ref{fig:RL}l and \ref{fig:RL}m. The total thickness is not much changed (Fig.\,\ref{fig:RL}n). The total number of needed iterations is substantially reduced compared to the Bayesian optimization only, and the randomly sampled devices can be deterministically optimized through the trained RL model. Moreover, the cascaded approach leads to larger final averaged reward and better optimized device performance than those obtained through the DQNN-only model, because the partially optimized Bayesian optimizer has helped filter out many bad devices and provided better initialized devices for the DQNN model. 

\begin{figure}
    \includegraphics[width=0.7\textwidth]{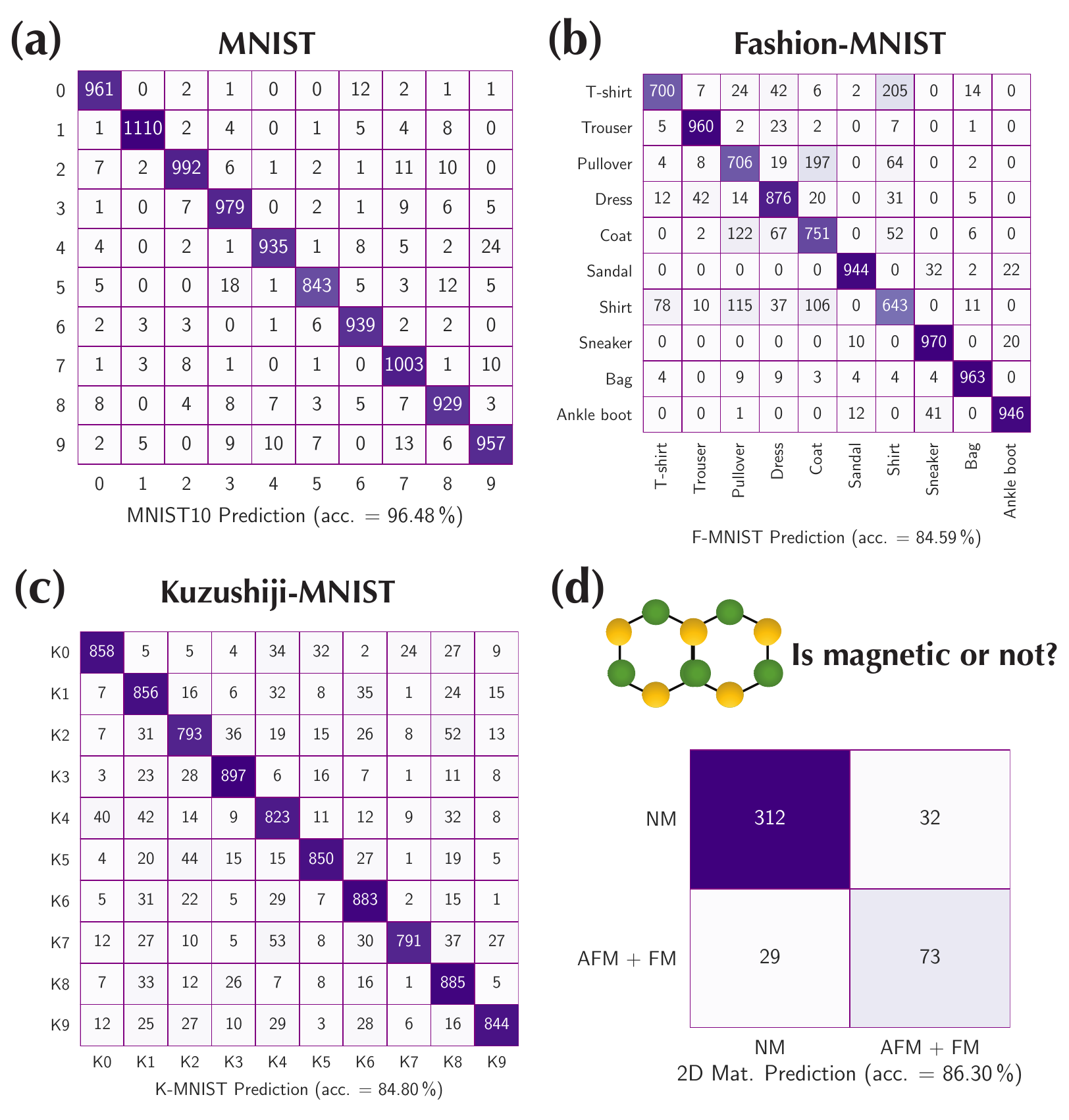}
    \caption{\label{fig:c_matrix} \textbf{Classification of images and magnetic nanomaterials.} The confusion matrices of image classification in the (a)\,MNIST dataset, (b)\,Fashion-MNIST dataset, (c)\,Kuzushiji-MNIST dataset. The confusion matrix of predicting magnetic and non-magnetic two-dimensional materials in the C2DB database\cite{HaastrupEtAl2018M}.}
\end{figure}

A general-purpose O-GEMM hardware can be deployed to accelerate a variety of ML applications. Figure\,\ref{fig:c_matrix} summarizes the deployment demonstration of the O-GEMM hardware with optimized devices in the tasks of the image classification in the MNIST, Fashion-MNIST (F-MNIST) and Kuzushiji-MNIST (K-MNIST) datasets, as well as the discovery of 2D magnetic nanomaterials in the ``Computational 2D Materials Database'' (C2DB) dataset\cite{HaastrupEtAl2018M}. We use multilayer perceptron (MLP) neural networks in these tasks and the calculations of all linear layers are performed through the O-GEMM hardware emulator; see \emph{Supplementary Information Sections 3 and 4} for the details of MLP architectures. Figure\,\ref{fig:c_matrix}a displays the confusion matrix of the classification of handwritten digits in the MNIST dataset. In order to have the best hardware deployment, we adopt a physics-aware training approach of executing the GEMM operations in the calculations of both backpropagated gradients and forward functions on the O-GEMM hardware emulator, so that all noises and quantization errors are incorporated in both training and inference processes. The obtained prediction accuracy is $96.48$\,\%. In addition, we compare the prediction accuracies obtained from different approaches of performing GEMM operations in linear layers, including the training and inference using general purpose graphics processor units (GPUs), GPU training and O-GEMM inference, as well as GPU training, O-GEMM fine training and O-GEMM inference. The approach of using GPU training and inference with nearly infinite bit accuracy yields a high accuracy $97.17$\,\%, which is close to the result obtained from physics-aware training. A large drop in prediction accuracy ($92.46$\,\%) is observed in the approach of GPU training and O-GEMM inference, because of the finite-bit-accuracy ($\sim5$\,bit) and random detector noise in the O-GEMM hardware. Despite the high accuracy of physics-aware training, the execution on O-GEMM hardware emulator is slow. We further create a hybrid two-step physics-aware training approach involving GPU training and O-GEMM fine training. Specifically, we start the training using GPU-implemented linear layers and replace trained linear layers with O-GEMM-implemented ones to continue the training process; see \emph{Supplementary Information Sections\,3 and 4} for details. This hybrid approach can substantially reduce training time, while maintaining nearly the same prediction accuracy ($96.48$\,\%) with that obtained from full O-GEMM training. The detailed comparison of prediction accuracy obtained from different approaches is shown in Table\,\ref{table:phy_aware_training}. 

The physics-aware training is particularly crucial when the errors from noises and quantization play an important role. As described in \emph{Supplementary Information Section\,2}, the detector signal-to-noise ratio decreases when incident light power decreases and the detector speed increases. In addition to the input power $100\,$mW and $1\,$GHz detector bandwidth used for RL-assisted optimization, \emph{Supplementary Table\,1} lists the prediction accuracies under different input power and detector bandwidth. The direct inference of GPU-trained models using the O-GEMM hardware leads to substantial accuracy drop with reduced input power and increased detector bandwidth at $10\,$GHz. Moreover, the accuracy drop as a function of input power is quicker with $10\,$GHz detectors than that with $1\,$GHz detectors. In contrast, the physics-aware training approach substantially boosts the prediction accuracies in both cases, highlighting its benefit in hardware deployment. 

In addition to the MNIST dataset, Figs.\,\ref{fig:c_matrix}b and c display the confusion matrices of the image classification in the F-MNIST and K-MNIST datasets. The prediction accuracies of F-MNIST and K-MNIST obtained from physics-aware training are $84.59$\,\% and $84.80$\,\%, which are close to the accuracies obtained from the hybrid training approach. Furthermore, we also apply the O-GEMM hardware emulator to a MLP model for predicting 2D materials magnetic property in the C2DB library, which is calculated using density functional theory (DFT)\cite{HaastrupEtAl2018M}. The one-hot encoded input features are limited to structural information of materials and we explicitly exclude any features calculated from DFT. The output labels from the C2DB dataset are ``non-magnetic (NM)'', ``antiferromagnetic (AFM)'', and ``ferromagnetic (FM)''. Since the number of AFM and FM materials is small, we group these two classes as one class ``AFM + FM'' and denote the meaning of this class as magnetic materials. As a result, the fast-executed MLP model can replace the time-consuming DFT for a quick prediction of material properties. The prediction accuracies obtained from physics-aware training and the hybrid training approach are $86.09$\,\% and $86.10$\,\%, which are both close to the GPU implementation with $86.10$\,\% accuracy. 

\begin{figure}
    \includegraphics[width=0.9\textwidth]{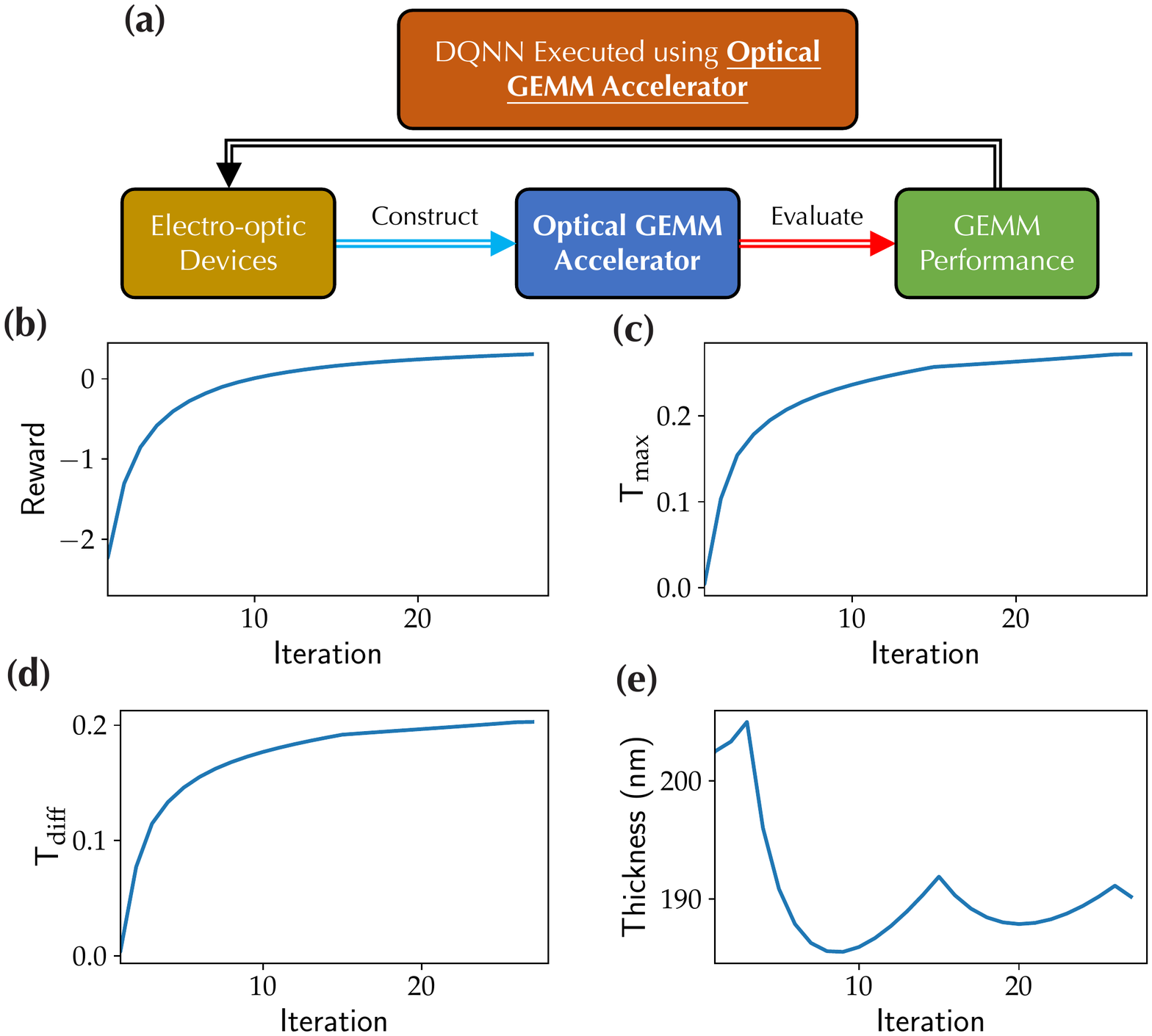}
    \caption{\label{fig:phcoreRL} \textbf{Optical general matrix multiplication (O-GEMM) accelerated design of optical machine learning accelerators.} (a)\,Diagram of employing O-GEMM hardware emulator to execute operations in linear layers of the deep Q-learning neural network (DQNN) model that is used to optimize the performance of reconfigurable unit. The accumulated average of (b)\,the reward calculated based on GEMM calculation accuracy, as well as (c)\,the maximum transmittance ($T_\mathrm{max}$), (d)\,the modulated transmittance ($T_\mathrm{diff}$), and (e)\,the total thickness of individual device as a function of iterations in the validation process using the DQNN model.}
\end{figure}

Finally, we deploy the O-GEMM accelerator emulator to execute GEMM operations in all linear layers in the DQNN model that is used to optimize the GST-based reconfigurable unit in O-GEMM accelerators. Thus, we close the design loop of O-GEMM hardware accelerator as shown in Fig.\,\ref{fig:phcoreRL}a. Figures\,\ref{fig:phcoreRL}b -- e summarize the reward, maximum transmittance, transmittance modulation, and total device thickness, as a function of iterations, which all show clear trends of optimizing GST-based reconfigurable unit in O-GEMM accelerators and the strong correlation between system calculation accuracy and device specifications.

\section*{Discussion}
\vspace{-20pt}

In contrast to the conventional ``bottom-up'' approach of designing optoelectronic devices without a close connection to the system performance requirement, our demonstrated holistic device-system co-design methodology provides a new ``top-down'' approach for an end-to-end optimization and offers insights into what device specifications are important to optimal system performance. Although the material and device parameter space in our demonstration is not gigantic, the developed highly-parallelized hardware emulator enables the further exploration of various RL algorithms in large-scale optimization problems. Moreover, the demonstrated physics-aware training approaches lay out strategies on how to deploy physical hardware systems to different ML application scenarios\cite{WrightEtAl2022N}.


\newpage
\begin{methods}

\subsection{Negative value representation.} In order to handle bipolar elements (negative and positive values), we represent both input vector $\vectorarrow{v}$ and weight matrix \textbf{W} as the difference of two positive vectors and matrices. Specifically, the MVM operation $\textbf{W}\vectorarrow{v}$ is calculated as $(\textbf{W}^{+} - \textbf{W}^{-})(\vectorarrow{v}^{+} - \vectorarrow{v}^{-})$, where $\textbf{W}^{\pm}$ and $\vectorarrow{v}^{\pm}$ contain only positive elements that can find corresponding physical quantities. The output vector $\vectorarrow{o}$ can be similarly decomposed as $\vectorarrow{o}^{+} - \vectorarrow{o}^{-}$, with $\vectorarrow{o}^{+} = \textbf{W}^{+}\vectorarrow{v}^{+} + \textbf{W}^{-}\vectorarrow{v}^{-}$ and $\vectorarrow{o}^{-} = \textbf{W}^{-}\vectorarrow{v}^{+} + \textbf{W}^{+}\vectorarrow{v}^{-}$. As a result, we need $4$ calls of the MVM calculation using optical hardware and one additional electronic subtraction to obtain the final output. 

\subsection{Multilayer thin film stack of PCM-based reconfigurable unit.} There are in total $6$ layers of thin films and we define the layer index as an integer in the range of $[0,5]$. Each layer has a film thickness in the range $[5,50]$\,nm. In each layer, the material can be selected from a material list containing Si$_3$N$_4$, Al, SiO$_2$, Au, ITO, and GST. We define the material index as an integer in the range of $[0,5]$ to represent the material following this order. For example, $0$ represents Si$_3$N$_4$ and $3$ represents Au. Each material corresponds to a complex-valued refractive index at $1.3\,\mu$m, which is obtained either from literature or from experimental measurements. The complex-valued refractive index of GST depends on the applied voltage. The complete list of materials complex-valued refractive indices is shown in \emph{Supplementary Table\,2}. A legitimate reconfigurable unit must contain active GST material, since it is the only material that can change optical properties to control light transmittance. In addition, considering the practical implementation, the adjacent layers of a GST-PCM layer need to be ITO material, since electric voltages or currents have to be applied through ITO to induce the phase transition in PCM materials and thus the change of refractive index. Based on all information, we define our device stack as a $10$-dimensional space and the details for each dimension are shown in \emph{Supplementary Table\,3}.

\subsection{Reward and action in reinforcement learning (RL) algorithms.} The reward in RL algorithms is defined in system level. Specifically, we randomly generate $10000$ pairs of input matrices with each element in the range $[-1,1]$. With a given selection of materials in each layer of devices and their corresponding thicknesses, the hardware emulator calculates the output vector with generated input matrices. Simultaneously, a PyTorch \texttt{matmul} function is also used to calculate the output vector, which is considered as the accurate value. We denote one element from the O-GEMM-calculated output vector as $\hat{v_o}$ and one element from the PyTorch-calculated output vector as $v_o$. The $i-$th calculation error $E_i$ is expressed as $E_i = \hat{v_o}_{,i} - v_{o,i}$. The reward is expressed as $1-10\times\mathrm{Std}_{i \in [1, 10000]}E_i$. $\mathrm{Std}$ represents the standard deviation of calculation errors of all $10000$ calculations. 
Regarding actions, we define a $20$-dimensional space and the action is a step-wise update of the previous device space. Essentially, each action dimension is the plus and minus step for the device space. The detailed description of the action shown in \emph{Supplementary Table\,4}.

\subsection{Thin film deposition and characterization.} GST-PCM and ITO materials were deposited through a Denton Discovery 18 sputtering system. The base pressure was set to be $<2\times 10^{-6}$\,Torr. The GST material was deposited through RF sputtering with a sputtering pressure $4.5\,$mTorr, a sputtering power $30$\,W, and an Argon gas flow $\sim100\,$sccm. The deposition speed for PCM was $\sim7\,$nm/min and a total thickness of $\sim10\,$nm was deposited. The structural change of GST from an amorphous phase to a crystalline phase was achieved by heating the film at $\sim200^{\circ}$C for $30\,$min. The ITO material was deposited through DC sputtering with a sputtering pressure $4.5\,$mTorr, a sputtering power $25$\,W, and an Argon gas flow $\sim100\,$sccm. The deposition speed for ITO was $\sim7.7\,$nm/min. The refractive indices of GST-PCM, SiO$_2$, and Si$_3$N$_4$ were measured using a VASE ellipsometer from J.\,A.\,Woollam in a range from $300\,$nm to $3\,\mu$m. The transmittance of fabricated thin film stacks were measured using a home-made spectrometer built from a PyLoN InGaAs 1D array and a SpectraPro HRS spectrograph, which are both from Teledyne Princeton Instruments. 

\subsection{Deep Q-learning neural network (DQNN).} 
The DQNN model consists of a $6$ dense-layer network that is trained as the Q table to generate appropriate actions. The  numbers of neurons propagating from the input to the output are $10$, $512$, $1024$, $512$, $256$, and $20$, respectively. A \texttt{tanh} activation is added between dense layers. The input is the current device structure. The outputs include a $10$-dimensional device information table (see \emph{Supplementary Table\,3}) and the Q value of a $20$-dimensional action table (see \emph{Supplementary Table\,4}). After a \texttt{softmax} function, we select the action that yields the largest Q value (reward) based on the current device structure. For the training process, we first randomly generate $2000$ states and action samples, and then store them along with the evaluated rewards into the memory. After that, we update the DQNN model using the data from the memory. The batch size we use is $128$ and the optimizer is \texttt{Adam} with the learning rate of $0.005$. The data in the memory is continuously updated through the training process. Epsilon-greedy is applied to balance the exploration and exploitation of action selection. The initial epsilon is set as $0.5$, with a decline rate of $0.04$. Once it drops to $0.1$ it remains at this level. $1000$ iterations are performed per epoch in the training process, and $500$ iterations are performed per epoch in the validation process. The device information is reinitialized at the beginning of each epoch. In the validation process, the randomly generated initial devices are selected to have reward $<0$. 

\subsection{Bayesian optimization.} For Bayesian optimization, the optimization space and corresponding bounds are the same as shown in \emph{Supplementary Table\,3}. The GEMM simulation is embedded directly to the Bayesian optimizer to calculate reward. 
There is no random exploration process and the optimization starts from the first iteration. Considering the fabrication limit, the thickness of the layers is rounded to the nearest integer before passing to the GEMM hardware emulator.

\subsection{Cascaded Bayesian optimizer and DQNN.} In the approach of cascading Bayesian optimizer and DQNN, $200$ iterations of Bayesian optimization are performed first. We then use the partially optimized Bayesian output as the initial device input of the DQNN, as shown in Fig.\,\ref{fig:RL}a. Here, the parameters of the Bayesian optimizer and DQNN model are the same as in the previous section. In the validation process, the generated initial devices from the Bayesian optimizer are also selected to have reward $<0$.

\end{methods}

\pagebreak
\begin{addendum}
\item [Data availability] Upon publication, all data that support the plots within this paper and other findings of this study will be available on a public \emph{GitHub} repository.  

\item [Code availability] Upon publication, all codes that support the plots within this paper and other findings of this study will be available on a public \emph{GitHub} repository. 
    
\item [Acknowledgements] We thank Dr.\,Kristian Sommer Thygesen for kindly providing us with the C2DB database. R.\,C. and W.\,G. thank the support from the University of Utah start-up fund. P.\,T.\,Z. and C.\,Y. thank the support from grants NSF-2019336 and NSF-2008144. 

\item [Author contributions] C.\,Y. and W.\,G. conceived the idea and designed the project. Y.\,T. performed machine learning algorithms and P.\,T.\,Z. optimized the hardware emulator under the guidance of C.\,Y. and W.\,G. R.\,C. experimentally fabricated and characterized materials and devices under the supervision of W.\,G. Y.\,T., C.\,Y., and W.\,G. wrote the manuscript. All authors discussed the manuscripts and provided the feedback. 

\item[Competing Interests] The authors declare that they have no competing financial interests.

\item[Correspondence] Correspondence and requests for materials should be addressed to Cunxi Yu (email: cunxi.yu@utah.edu) and Weilu Gao (email: weilu.gao@utah.edu).
\end{addendum}

\newpage

\begin{figure}
    \includegraphics[width=0.9\textwidth]{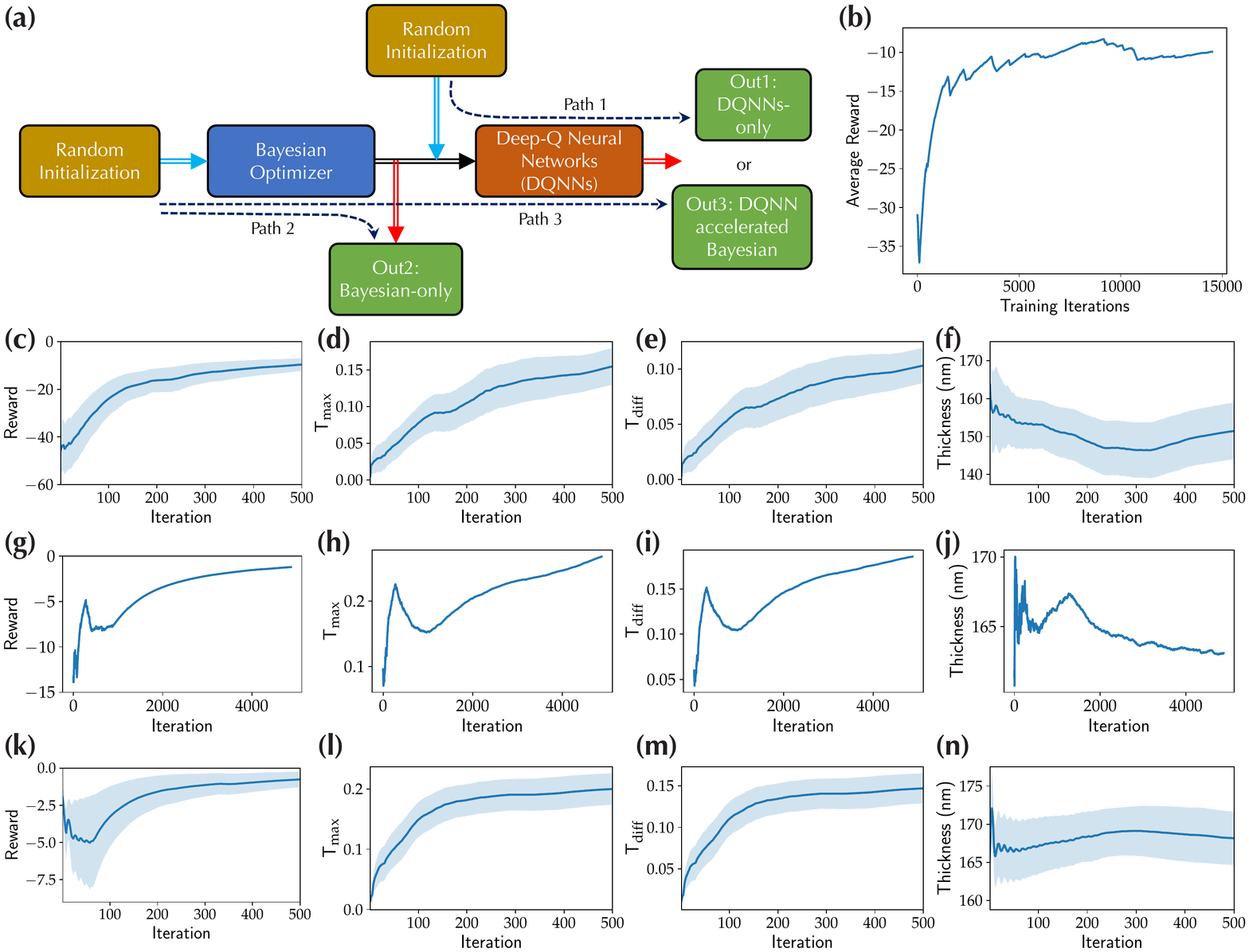}
    \caption{\label{fig:RL} \textbf{Reinforcement-learning-assisted design of phase-change-material-based spatially reconfigurable array.} (a)\,Diagram of three employed methods. The first path is based on deep Q-learning neural network (DQNN) with random initialization. The second path is based on Bayesian optimization with random initialization. The third path is a method of cascaded Bayesian optimization and DQNN. The initialization of DQNN is provided by a small-iteration-run Bayesian optimizer. (b)\,An representative training curve of the DQNN model. The accumulated average of (c)\,the reward calculated based on GEMM calculation accuracy, as well as (d)\,the maximum transmittance ($T_\mathrm{max}$), (e)\,the modulated transmittance ($T_\mathrm{diff}$), and (f)\,the total device thickness as a function of iterations in the validation process using the DQNN model. (g) -- (j)\,The same quantities obtained in the validation process using Bayesian optimization. (k) -- (n)\,The same quantities obtained in the validation process using cascaded Bayesian optimization and DQNN model.}
  \end{figure}

\newpage
\bibliography{/Users/weilugao/Dropbox/Utah/Research/bib_global/weilu.bib} 

\newpage

\begin{table}
  \centering
  \caption{The prediction accuracy comparison of different training and inference methods}
  \label{table:phy_aware_training}
  \begin{tabularx}{\textwidth}{|| X || X | X | X | X ||}
    \hline
    Dataset & GPU train + GPU inference & GPU train + O-GEMM inference & GPU train + O-GEMM fine train + O-GEMM inference (hybrid training approach) & O-GEMM train + O-GEMM inference (physics-aware training) \\ [0.5ex] 
    \hline\hline
    MNIST & {97.17\,\%} & {92.46\,\%} & {96.53\,\%} & {96.48\,\%}\\
    \hline
    F-MNIST & {87.47\,\%} & {82.26\,\%} & {84.67\,\%} & {84.59\,\%}\\
    \hline
    K-MNIST & {83.77\,\%} & {79.23\,\%} & {84.79\,\%} & {84.80\,\%}\\
    \hline
    Materials Discovery & {86.10\,\%} & {83.00\,\%} & {86.09\,\%}  & {86.10\,\%}\\
    \hline
    \hline
  \end{tabularx}
  \\
\end{table}


\end{document}